# M Giant Kinematics in Off-Axis Fields
# Between 150 and 300 Parsecs
# from the Galactic Center

R. D. Blum[1,2], J. S. Carr[2], K. Sellgren[1,3], and D. M. Terndrup[4]

Department of Astronomy, The Ohio State University

174 W. 18th Ave., Columbus, Oh, 43210

## ABSTRACT

We present radial velocities for approximately 40 stars in each of three optically obscured, off-axis, fields toward the Galactic bulge. Combined with the data presented by Blum et al. (1994), we now have mean radial velocities and radial velocity dispersions in four fields towards the Galactic bulge. These four fields lie nearly along an axis whose position angle from the major axis of the Galaxy is 55°. The observed kinematics generally match both Kent's (1992) axisymmetric and Zhao's (1994) barred dynamical model predictions, but are marginally better described by the latter.

The velocity dispersion in our innermost field is high, $153 \pm 17$ km s$^{-1}$. Our data, combined with that from previous studies at larger radii, suggest that the stellar velocity dispersion is flat or still rising at projected radius R $\lesssim 150$ pc.

*Subject headings:* Galaxy: center — Galaxy: kinematics and dynamics — stars: giants — techniques: radial velocities

## 1. INTRODUCTION

Recent photometric studies and star count data have produced strong evidence that the inner Galaxy, or bulge, is bar shaped (Blitz & Spergel 1991, Nakada et al. 1991, Weinberg 1992, Whitelock & Catchpole 1992, Weiland et al. 1994, Stanek et al. 1994, and Dwek et al. 1994). Dynamical modeling of the observed gas kinematics also indicates that the inner Galaxy is non-axisymmetric (Binney et al. 1991), a result first suggested by de Vaucouleurs (1964), who compared the non-circular gas motions of the inner Galaxy to those in external barred galaxies.

Surprisingly, clear signs of a non-axisymmetric potential have not been found in the observed stellar kinematics (de Zeeuw 1993). Recent stellar kinematic studies have been completed at a number of positions projected onto the Galactic bulge. The majority of these have been in the optical windows, regions of lower interstellar extinction, along the bulge minor axis ($b \gtrsim 3°$), and several have explored the bulge kinematics at larger longitudes (e.g. Minniti et al. 1992 and Harding & Morrison 1993). Kent (1992) and de Zeeuw (1993) provide recent compilations of the locations and results of these bulge kinematic studies. At radii inside the optical windows, the majority of stellar kinematic work has been done at the Galactic center (GC) (R $\lesssim$10 pc; see Genzel, Hollenbach, & Townes 1994 for a recent review) and along the major axis between $\sim 10 - 100$ pc (Lindquist et al. 1992). Our data, combined with that of Blum et al. (1994), explore the inner Galaxy stellar kinematics at radii which are intermediate between the Galactic center and optical windows. To search for evidence of triaxiality and to better constrain the inner Galaxy mass distribution at these intermediate radii, we have obtained radial velocities for stars in each of four fields located at off-axis positions between 150 and 300 pc projected radius from the GC. The results for one of these fields were previously given by Blum et al. (1994).

The large amount of interstellar extinction toward the GC at optical wavelengths requires spectroscopic observations at near infrared or longer wavelengths. The radial velocities presented here were obtained from spectra centered near the 2.3 $\mu$m CO bandhead, a strong photospheric absorption feature in M giants. Where required, we adopt a distance of 8 kpc to the GC (Reid 1993).

## 2. OBSERVATIONS AND DATA REDUCTION

The observational data are a combination of three distinct sets: $J$ and $K$ band images from which the program stars were selected, spectra obtained with the CTIO grating spectrometer, IRS, and spectra obtained with the IRTF echelle spectrometer, CSHELL.

These three data sets and their reduction procedures are discussed separately below. For all three data sets, the basic image processing tasks were performed using IRAF.[5]

## 2.1. Imaging

The program stars are located in four fields which lie along a "slit" projected across the bulge (approximately intersecting $l = b = 0°$) and oriented at an angle of $\sim 55°$ with respect to the bulge major axis. The field locations were chosen to coincide with foreground SAO stars so that accurate offsets could be calculated to the program stars for use in obtaining the program star spectra (see below). The locations of all four fields are indicated in Table 1.

The program stars were selected from the brighter and redder stars on $J$ ($\lambda$=1.25 $\mu$m, $\Delta\lambda$=0.24 $\mu$m) and $K$ ($\lambda$=2.2 $\mu$m, $\Delta\lambda$=0.4 $\mu$m) band images obtained with the Ohio State Infrared Imaging System (OSIRIS) on the Perkins 1.8m telescope near Flagstaff, Arizona in 1993 May. OSIRIS employs a 256$\times$256 NICMOS III array. The images were taken in low resolution mode which results in a spatial scale of $\sim 1.5''$ pix$^{-1}$ on the Perkins telescope. OSIRIS is more fully described by DePoy et al. (1993).

All of the $J$ and $K$ images were taken through a 10 % transmission neutral density filter so that the brighter stars would not saturate the detector. A typical exposure time with the neutral density filter was 3 seconds for each filter. Each $J$ and $K$ image was sky subtracted with frames obtained at nearby positions off the bulge, then flattened using dome flats. Several of the sky images used were taken without the 10% transmission neutral density filter. These were scaled appropriately before subtraction. Low order variations due to scattered light differences may exist between images with and without the neutral density filter, but any residual background should be accounted for by local background values determined when the photometry was obtained for the stellar images (see below).

$J$ and $K$ magnitudes were obtained from the reduced images by synthesizing circular apertures (3 pixel radius) about each star. A local background, or sky, component (determined from an annulus extending 5 pixels from the aperture edge) was subtracted before the magnitude determination. Magnitudes were typically obtained for each program star from three frames each at $J$ and $K$. Scatter between measurements of the same star from the different frames suggests typical photometric uncertainties of $\pm$ 0.05 mag and $\pm$ 0.02 mag at $J$ and $K$, respectively.

---

[5]IRAF is distributed by the National Optical Astronomy Observatories.

The photometry was calibrated using stars of known magnitudes from Baade's window ($l$, $b$ = 1°, −4°) (stars 107, 108, B 28, and B 66 from Frogel and Whitford 1987; hereafter FW) and the flux standards HD 162208, HD 106965, and HD 136754 (Elias et al. 1982). These stars have magnitudes given on the CTIO/CIT system. The scatter between different measurements (typically 3 to 5 frames for each filter) of the same flux standard was $\lesssim$0.03 mag and 0.02 mag at $J$ and $K$, respectively. There is an additional uncertainty of $\sim \pm 0.04$ mag due to uncertain airmass correction between the standards and program stars.

Adding the above photometric uncertainties in quadrature results in a typical uncertainty for an individual program star of 0.07 mag and 0.05 mag at $J$ and $K$, respectively. The true uncertainty may be larger than this due to image crowding in the field. However, the effects of crowding should be less for the brighter stars which we are concerned with here.

Transformations between the OSIRIS photometric system and other systems do not yet exist, but the color transformation between the standard OSIRIS system (no neutral density filter) and the CTIO/CIT system is estimated to be less than the observational uncertainties reported here and is not included (Blum et al. 1994; Tiede & Frogel 1994). However, there may be a systematic color correction for the neutral density filters. Sufficient data are not yet available for OSIRIS to determine whether a transformation exists between the standard and neutral density filter systems.

The observed $K$ and $J − K$ for each star in fields 1, 2, and 3, for which spectra were obtained, are given in Table 2.

## 2.2. CTIO Spectra

The spectroscopic observations for fields 1 and 3 were made on the CTIO 1.5m telescope in 1993 July using the facility infrared spectrometer (IRS) which, at that time, employed an SBRC 58×62 InSb detector. The spectral resolution was 84 km s$^{-1}$ (42 km s$^{-1}$ pix$^{-1}$) at 2.3 $\mu$m, the position of the CO bandhead. The IRS had a spatial scale of 2.39$''$ pix$^{-1}$ and a 4.8$'' \times 30''$ N–S oriented slit. A detailed description of the IRS may be found in DePoy et al. (1990).

The instrument set up and data taking mode were the same as that for field 4, as described in Blum et al. (1994). Briefly, program stars were acquired by offsetting from a visible star in the field using calculated offsets (Table 2) from a $K$ band image. The visible stars from which program stars were located for fields 1 and 3 were SAO 185604 ($\alpha$(1950)=17$^h$ 37$^m$ 13.48$^s$, $\delta$(1950)=−28° 53$'$ 50.0$''$) and SAO 185927 ($\alpha$(1950)=17$^h$ 51$^m$

$48.83^s$, $\delta(1950)=-28°$ 44' 53.7''), respectively. After offsetting, spectra were obtained in a star–sky–sky–star sequence.

The data reduction for these spectra was identical to that described by Blum et al. (1994), including wavelength calibration and correction for telluric absorption features, with one minor exception. Instead of combining both star and sky frames from a given star–sky–sky–star sequence resulting in a single spectrum for each program star, only the sky frames were combined. Each star frame was reduced and analyzed separately. This allowed for a first order check on the accuracy of the radial velocities by directly comparing the values from the two spectra.

### 2.3.  IRTF Spectra

The spectra for stars in field 2 were obtained in July 1994 using the IRTF 3m infrared telescope and facility echelle spectrometer, CSHELL (Tokunaga et al. 1990 and Greene et al. 1993). CSHELL employs a $256\times256$ InSb array and has a plate scale of 0.2'' pix$^{-1}$. The slit width was set at 2'', giving a spectral resolution of 25 km s$^{-1}$ (2.5 km s$^{-1}$ pix$^{-1}$). The 30'' long slit was oriented E–W.

Field 2 was centered on SAO 185837 ($\alpha(1950)=17^h$ $47^m$ $53.3^s$, $\delta(1950)=-28°$ $38'46.8''$). CSHELL has a flip–in mirror which allows for imaging the field without moving the grating. This mirror was used to center each program star in the slit after offsetting from SAO 185837. Corrections made to the initial offsets for program stars suggest that the offsets given in Table 2 are accurate to about $1'' - 2''$. The higher spatial resolution of CSHELL allowed for a more efficient data taking mode since crowding is less at this scale. Each program star was observed in two positions, $\sim 7'' - 8''$ E and $\sim 7'' - 8''$ W of the slit center. In this way, star and sky measurements were made with two instead of four frames.

The final spectrum for a typical program star was obtained in the following way. The initial images were flat–fielded using continuum lamp images. Bad pixels were linearly interpolated across and replaced. The resulting two images for each star were then subtracted from each other. Two spectra were then extracted from the subtracted image using the IRAF "apextract" package. Each one-dimensional spectrum resulted from synthesizing 10 pixel wide apertures which had been traced along the spectral dimension and which included a background aperture on each side of the spectrum located 13 pixels from the spectrum center. These spectra were ratioed by the spectrum of a hot star (spectral types O, B, or A) chosen from the Bright Star catalog to cancel telluric absorption features. Finally, the spectra were wavelength calibrated using Ar and Kr lamp lines.

The wavelength response changes slightly along the spatial dimension of the CSHELL array. Therefore, hot stars, radial velocity standards, and wavelength calibration images were taken at both of the spatial positions corresponding to the program stars. When determining velocities (see below), the IRTF/CSHELL data were always compared to standards taken at the same spatial position on the array.

Figure 1 shows the spectra of a radial velocity standard and two program stars measured on the CTIO/IRS and IRTF/CSHELL systems.

## 3. SPECTRAL ANALYSIS

### 3.1. Radial Velocities

Radial velocities were obtained for the program stars in all three fields by cross correlation with stars of known velocity using the "RV" package in IRAF. Typically all three, but at least two, of the following stars from the Bright Star catalog were observed each night on both the CTIO/IRS and IRTF/CSHELL systems: BS 6459 (K5 III), BS 7323 (M0 III), and BS 5192 (M5 III). In addition, we observed the velocity standard used by Blum et al. (1994), star 301 from Sharples, Walker, & Cropper (1990).

No systematic changes in velocity were observed for a given standard taken at different times during a given night. Typical differences in multiple measurements of the same standard were less than 10 km s$^{-1}$, so all spectra of a given standard were averaged together on each night. Cross correlating the different standards on each night resulted in measured differences between the heliocentric velocities of less than 10 km s$^{-1}$ compared to the differences in the published values (including star 301) in all cases except one, in which the difference between BS 6459 and BS 7323 on one night was 14 km s$^{-1}$.

For all nights, two standards (of the 2 or 3 available) were chosen to cross correlate with the program stars. In each case standards were used whose measured heliocentric velocity difference compared to the difference in the published values was less than 5 km s$^{-1}$. For one night on the CTIO/IRS system, this included star 301. Combined with the two measurements of each program star, this procedure resulted in four measured velocities for a typical program star. We adopt the average of these (typically) four velocities as the velocity for a given program star (Table 2). The standard deviation for a program star velocity calculated in this way was less than 10 km s$^{-1}$ for 95 % of the program stars; the maximum deviation was 21 km s$^{-1}$.

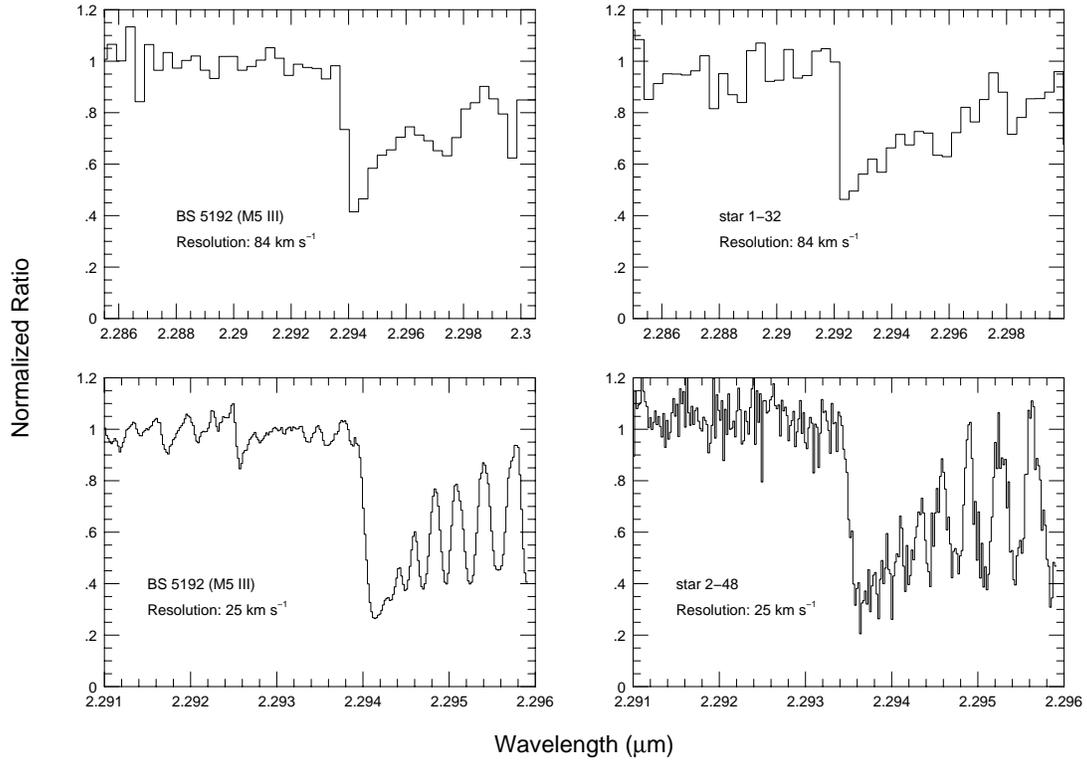

Fig. 1.— Spectra of the 2.3 $\mu$m CO bandhead. Upper panels, CTIO IRS spectra: resolution = 84 km s$^{-1}$ (42 km s$^{-1}$ pix$^{-1}$). Lower panels, IRTF CSHELL spectra: resolution = 25 km s$^{-1}$ (2.5 km s$^{-1}$ pix$^{-1}$). A typical radial velocity standard (BS 5192, M5 III) is shown along with program stars 1−32 and 2−48.

A small number of stars measured on the IRTF/CSHELL system had radial velocities large enough that the bandhead was nearly shifted off the array. For these stars (indicated in Table 2), the velocity was estimated by measuring the shift in the bandhead position relative to the bandhead position for the standard. Because the CSHELL resolution is quite high (25 km s$^{-1}$) and a resolution element is well sampled (10 pixels), this should give a velocity estimate within the quoted uncertainty (see below). Comparing the velocity measured in this way and that measured by cross–correlation for a number of stars indicates that this is the case.

An overall uncertainty in the measurement of a program star velocity was estimated by comparing the velocities determined for a subset of program stars on both the CTIO/IRS and IRTF/CSHELL systems. The difference between velocities measured for 15 stars in all four fields on both systems was 12 km s$^{-1}$ ± 15 km s$^{-1}$, in the sense that the CTIO/IRS velocities are more negative. We adopt a typical error of ± 15 km s$^{-1}$ for the measurement of a single program star velocity. We discuss the effects of a possible systematic offset below.

## 3.2. CO Strengths

Measurement of the strength of the 2.3 $\mu$m CO bandhead leads to an estimate of the intrinsic, or de-reddened, color of the program stars. This combined with the observed $J - K$ color yields the reddening to each field. The CO feature is sensitive to both temperature and gravity, becoming stronger in lower gravity and cooler stars (Baldwin et al. 1973; Kleinmann & Hall 1986; hereafter KH). CO strengths in excess of 20 % (as measured by KH and here; see below) are characteristic of late K and M giants and supergiants. The difference in strength between dwarf and giant M stars is large, so they are easily distinguished from each other. The similar effects of lower gravity and lower temperature which both result in increased CO strength mean that K supergiants could have as strong a CO strength as later M giants. The more luminous K supergiants, located near or in front of the GC, would be conspicuous by their brightness and color and would not have been selected (see below). It is possible that K supergiants, located behind the GC, could be confused with M giants closer to the GC. However, luminosity class I stars are expected to be much more rare, so we assume that the program stars are giants.

We define the measured CO strength as $1 - F_{CO}/F_{con}$, where $F_{CO}$ and $F_{con}$ are the total flux in bands just longward and shortward of the bandhead, respectively. For the CTIO/IRS data, the band passes were 0.005 $\mu$m wide. This value was chosen so a comparison could be made with the giant stars of known spectral type in KH. For the IRTF/CSHELL data, the total wavelength coverage was small ($\approx 0.005$ $\mu$m) and a direct comparison could not be made in this way. For these stars, the CO strength was measured in band passes 0.001 $\mu$m wide. The CO strengths (Table 2) were calculated using the LINER spectral analysis program in use at Ohio State (written by Richard Pogge).

The stars measured with the CSHELL/IRTF system (field 2) were first converted from a measured CSHELL/IRTF CO strength to a CTIO/IRS one based on the measurements of the subset of stars which were measured on each system (16 stars including the radial velocity standards). A linear fit to the data suggest a CO strength for field 2 of $0.29 \pm 0.03$. The large uncertainty results from the considerable scatter in the measured CO strengths. About $^{1}/_{2}$ of these stars were taken under cloudy conditions, and their signal–to–noise ratio is much lower. Fitting the higher signal-to-noise data only, suggests a CO strength of $0.32 \pm 0.02$. We adopt this value for the CTIO/IRS CO strength of field 2.

The mean CO strengths in fields 1,2, and 3 were found to be, respectively, $0.30 \pm 0.01$, $0.32 \pm 0.02$, and $0.29 \pm 0.01$, where the quoted uncertainty is the uncertainty in the mean (except for field 2). The CO strengths in fields 1, 2, and 3 are similar to that found for field 4 $(0.33 \pm 0.01)$ by Blum et al. (1994).

The measured CO strengths for the velocity standards on the CTIO/IRS system can also be compared to the CO strengths for stars of the same spectral type given in KH. The KH spectral resolution and band passes are similar to those used here for the CTIO/IRS data, so a direct comparison can be made. The CO strengths are in excellent agreement for BS 6459 $(0.23 \pm .01,$ K5 III) and BS 7323 $(0.25 \pm 0.02,$ M0 III) and reasonable agreement for BS 5192 $(0.30 \pm 0.01,$ M5 III). We therefore conclude that no transformation is necessary between our data and KH.

The mean CO strength can be used to estimate the mean intrinsic $J - K$ color in each field. The difference between the observed and intrinsic color yields the color excess. The color excess and an assumed interstellar extinction law lead directly to the extinction $(A_K)$ for each field. At least two factors could limit the accuracy of this approach. First, the relationship between CO strength and intrinsic $J - K$ has been found to be different for different stellar populations. FW measured photometric CO indices (a measure of the CO absorption strength over a 0.08 $\mu$m wide wavelength interval compared to a similar continuum band using narrow-band filters) for M giants in Baade's window and found that these stars had stronger indices (i.e. stronger CO absorption) than disk M giants of the same $J - K$. The relationship between CO and $J - K$ may also change systematically with latitude in the bulge (Frogel et al. 1990). The second limiting factor is the significant scatter in the measured CO index versus spectral type or $J - K$ for the latest type bulge and disk M giants (Frogel et al. 1978, FW).

To estimate the intrinsic $J - K$ in our fields from CO indices and $J - K$ for either the field giants (Frogel et al. 1978) or bulge giants (FW), we must first convert our spectroscopic CO strength to a CO index. KH showed that the CO index employed by Frogel et al. (1978) for disk M giants is strongly correlated with the spectroscopic CO strength. Assuming this same relationship holds for our stars, we can assign CO indices to the measured spectroscopic CO strengths and use the CO index and $J - K$ data for bulge giants (FW) and disk giants (Frogel et al. 1978) to estimate the intrinsic color for stars in our fields.

The intrinsic $J - K$ for each field was obtained by averaging the bulge (FW) and disk (Frogel et al. 1978) $J - K$ values over a range of CO indices which corresponded to the mean CO indices $\pm$ one standard deviation in each of our fields. The intrinsic $J - K$ and $A_K$ (determined with the interstellar extinction law of Mathis 1990) are similar for both the disk and bulge determinations. We adopt the bulge values to determine the unreddened color and $K$ magnitude in the program star fields. The CO index, mean intrinsic color, and $A_K$ for each field are given in Table 1.

### 3.3. Color Magnitude Diagram

In Figure 2, we plot the de-reddened color magnitude diagram (CMD) for all four fields using the intrinsic $J - K$ and $A_K$ estimated above from the bulge giant (FW) CO indices and $J - K$ values. The left panel shows the CMD for all stars with observed $K \lesssim 9$ mag. The right panel shows those stars for which we obtained spectra. There appears to be a clear sequence in the CMD, which we take to be the bulge or inner Galaxy giant branch. The relation for Baade's window giants is also plotted from the data in FW. The CMD for the stars which we observed spectroscopically, appears to have a slightly bluer component than the M giants in Baade's window. This could result from the uncertainties in the derived intrinsic $J - K$, a potential unknown color transformation (see §2.1), or from physical differences in the populations observed.

The data for Baade's window suggest that the CMD has larger width in $J - K$ at its bright end. Our data also show this trend, but seem to be even broader. Part of the spread in our data may be due to differential reddening within a given field and uncertain reddening corrections between fields. The stars in Figure 2 which contribute the most to the width in the CMD are from fields 1 and 2 which are located at lower latitude. The images of Glass, Catchpole, & Whitelock (1987) show that the extinction inside a few degrees of the GC is both strong and variable on small spatial scales. Our inner fields are more crowded than Baade's window which could also increase the scatter in our aperture photometry.

Some of the stars in the right panel of Figure 2 may be foreground or background stars. We have computed the distribution of stars along the line of sight for a simple two component bulge/disk model for each field with the same model described by Blum et al. (1994). The relative disk to bulge contamination for fields 1,2, and 3 (25%–29%) is quite similar to that reported for field 4 (28%) by Blum et al. (1994). In these simple models, which employ an exponential disk (Garwood & Jones 1987) and axisymmetric bulge (Kent 1992), many of the disk stars would still be at distances making them co-spatial with bulge stars. We also note that while there is substantial evidence for an asymmetric light distribution in the inner Galaxy, which argues against using this simple two component axisymmetric model, none of the non–axisymmetric models are fit to the near infrared light distribution inside $b \lesssim 3°$.

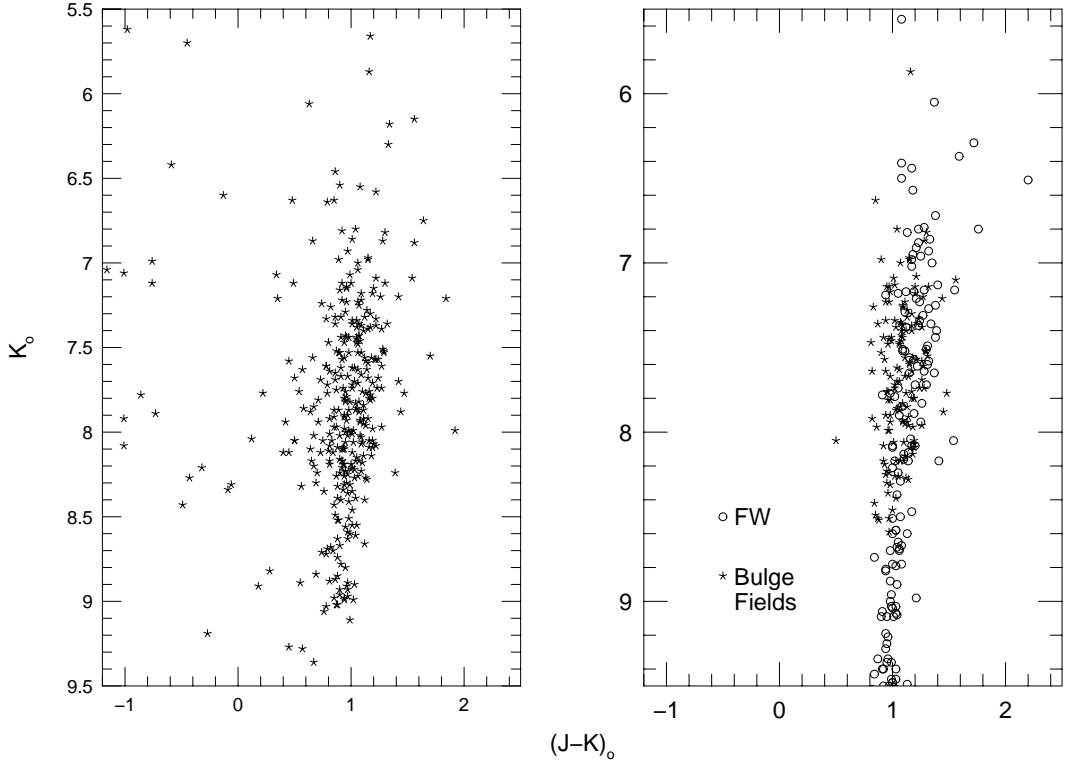

Fig. 2.— De-reddened color-magnitude diagram for the four program star fields. The extinction correction assumes the interstellar relation from Mathis (1990) and relies on using the mean CO strength to determine intrinsic color; see text and Table 1. The left panel depicts the CMD for stars in the four fields with observed $K \lesssim 9$ mag. The right panel shows the stars observed spectroscopically which were selected based on color and magnitude. The open circles in the right panel represent the data for Baade's window M giants taken from FW.

## 4. DISCUSSION

The mean heliocentric radial velocity and radial velocity dispersion for each field are given in Table 1. The quoted uncertainties are dominated by statistical uncertainty due to the small number of stars in each field, but they include the effects of measurement error (the uncertainties were determined as shown in Blum et al. 1994). These values may be compared to the predictions of dynamical models to help constrain the inner Galaxy mass distribution. In light of the photometric and gas kinematic observations, noted above in §1, we expect the inner Galaxy mass distribution may be strongly barred.

A range of bar models fit the observed near infrared light distribution about equally well (Dwek et al. 1994). These models have rather large differences in some of their characteristics, particularly in the orientation of the bar to our line of sight. To determine how sensitive the predicted kinematics are to the different mass models, we will compare our observed kinematics to predictions from two models which take quite different forms for the inner Galaxy mass distribution: the axisymmetric model of Kent (1992) and the triaxial bar model of Zhao (1994). The Kent model is an oblate, isotropic rotator. The velocity dispersion is assumed to be the same in any direction; the bulge appears flattened in this model due to net rotation about the axis normal to the plane of the Galaxy. Zhao's model is triaxial. It employs one of the best fit (Gaussian) density distributions found by Dwek et al. (1994) to the near infrared surface brightness distribution as observed by the DIRBE experiment on board COBE. The semi-major axes are 1.49:0.58:0.40 kpc (scaled from 8.5 kpc as adopted by Dwek et al. to 8 kpc). The major axis of the bar is oriented at an angle to our line of site of 13.4°. Both models assume a constant stellar mass to light ratio.

We note that the surface brightness asymmetry, which leads to the consideration of triaxial models for the bar in the first place, is constrained only with data at $b \gtrsim 3°$. Zhao (1994) employs an axisymmetric density component similar to Kent (1992) within this distance of the Galactic center. This means that Zhao's model is more similar to Kent's model in the region where the present observations are taken. However, even though the density distribution is similar, other characteristics, like the velocity dispersion isotropy are not.

Within the framework of its assumptions, the Kent (1992) model should make quite accurate predictions. The Zhao (1994) model is the most realistic attempt at modeling the Galactic bulge as a triaxial stellar system; however, more work will be needed to explore the effects of varying the bar parameters.

We discuss the observations and models within the context of the velocity dispersions, mean velocities, and velocity distributions. We conclude our discussion by considering new types of observations which will further establish the bulge mass distribution.

## 4.1.  Velocity Dispersion

Figure 3 shows a plot of our inner Galaxy dispersion measurements compared to the model results of Kent (1992) and Zhao (1994). The figure also shows observational results from the optical windows near the minor axis. The Kent prediction is for a "slit" projected across the bulge at an angle of $\sim 55°$ from the major axis. Kent generously made available his code to us so that these predictions could be made. The code is the same one that produced the Kent (1992) bulge model, with minor modifications. We reproduced the results of Kent (1992) to within a few km s$^{-1}$ at all radii with this code. The slit nearly encompasses all our positions. The model prediction changes by only a few km s$^{-1}$ at all radii for position angles from $45°$ to $90°$, so a single position angle is sufficient for comparison to the observations (ours and previous observations made on the minor axis). The bar model prediction for the minor axis was taken directly from a figure presented by Zhao (1994). The bar model reaches slightly higher dispersions and reaches its maximum dispersion at a smaller radius than the axisymmetric model. Both models generally fit our data.

It is interesting to note that the N body simulation of Sellwood (1993), which forms a bar through a disk instability, also predicts higher velocity dispersions in our fields ($140 - 150$ km s$^{-1}$, if his model is scaled to give a dispersion of 113 km s$^{-1}$ in Baade's window). This result is for the end state of Sellwood's simulation. We are grateful to Sellwood for providing us with the results of his simulation which we "observed" along lines of sight to our fields (the bar major axis was aligned at $\sim 20°$ to the line of sight). It is difficult to assess the significance of this result since the Sellwood model employs a rigid spherical mass distribution (30 % of the total mass) in addition to the distribution defined by the end state of the mass particles.

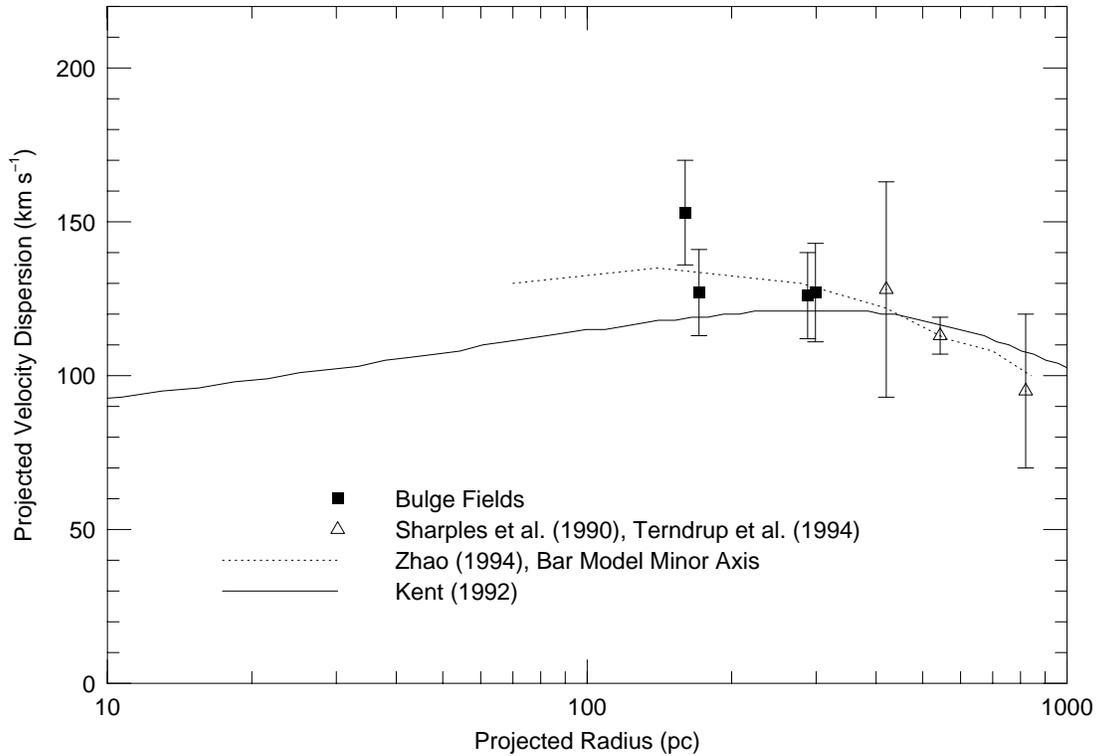

Fig. 3.— Comparison of observed and model projected velocity dispersions. The solid box data points are from the present paper, except the one at the largest radius which is taken from Blum et al. (1994). The data point at ∼ 500 pc (open triangle) is taken from the M giant study by Sharples et al. (1990) in Baade's window. The remaining two points (also open triangles) are from minor axis optical windows (Terndrup et al. 1994). The solid line represents the prediction from the Kent (1992) dynamical model (private communication; see text). This prediction is for a "slit" oriented at 55° which is similar to the distribution of our four fields (solid boxes); see text. The dashed line represents the minor axis prediction of Zhao (1994) as scaled from one of his figures.

Previous kinematic results for minor axis fields at larger $b$ (Rich 1990; Sharples et al. 1990; Terndrup, Frogel, & Wells 1994) and for fields at larger $l$ (Minniti et al. 1992; Harding & Morrison 1993) are also roughly consistent with the kinematics predicted by both the Kent (1992) and Zhao (1994) models (the mean radial velocity in Harding & Morrison's field is better matched by the Zhao predictions discussed below).

The dispersion in our innermost field is $153 \pm 17$ km s$^{-1}$, higher than either model prediction and higher than observed anywhere else in the inner Galaxy. This means that we have yet to observe the predicted velocity dispersion turnover and decrease toward smaller radii. Both models discussed here predict this, largely as a result of their adopted density profiles which are similar inside about 400 pc (minor axis distance). An average value for the dispersion in the central pc of the Galaxy is 125 km s$^{-1}$ (Sellgren et al. 1990) and in the central 10 pc may be as low as 75 km s$^{-1}$ (Rieke & Rieke 1988). If the bulge population joins onto the nuclear population, then we should expect to see the bulge dispersion turn over. We plan to obtain the kinematics for stars inside about 70 pc in an upcoming observing run to address this question.

Although not statistically significant (at least for the Zhao 1994 model), the inner field data point is 1 to 2 sigma above either model prediction. If the dispersion continues to rise or remain flat inside this position, the stellar mass to light ratio might not be constant. Kent (private communication) has made a preliminary check of the 3.5 $\mu$m and 4.8 $\mu$m COBE surface brightness distributions. These longer wavelength measurements should be less susceptible to uncertainties caused by interstellar extinction. While the spatial resolution is low (0.7° beam), Kent finds that the amount of light observed is consistent with his model (Kent 1992) and thus also with that of Zhao (1994). If the the velocity dispersion does not fall inside 100 pc, then a higher mass may be required, but the observed amount of near infrared light would remain unchanged.

Lindquist et al. (1992) have obtained velocities for OH/IR stars at projected radii between 10 and 100 pc. They find dispersions between about $50 - 100$ km s$^{-1}$ suggesting that the dispersion does fall in this range. However, the OH/IR stars also have considerable rotational velocities (up to 100 km s$^{-1}$ at 100 pc, much higher than the two models discussed here predict; see the next section) and so probably represent a different stellar population than the stars and models depicted in Figure 3.

## 4.2. Mean Radial Velocities

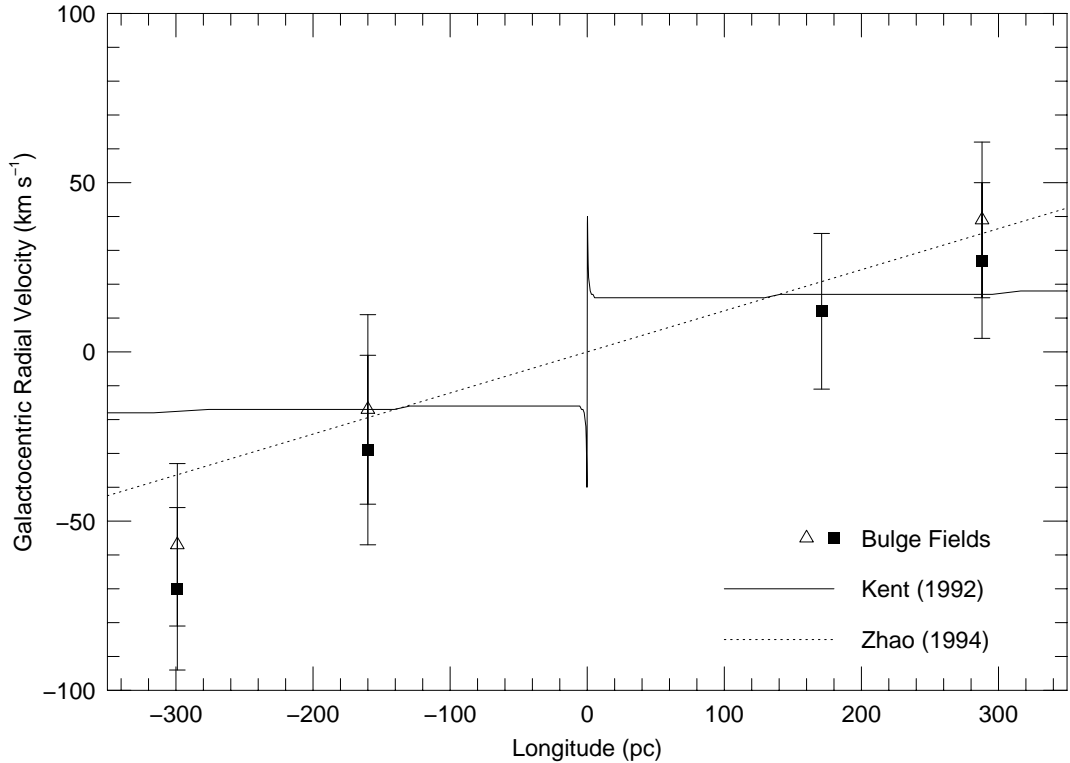

Fig. 4.— Comparison of observed and model mean velocities. All velocities in this figure are galactocentric. The solid boxes represent the four inner Galaxy fields discussed in this paper and Blum et al. (1994). The solid line is from Kent's (1992) model. The dashed line is an estimate from the Zhao (1994) model. The open triangles represent the effect of a possible systematic offset for the fields observed at CTIO based on a small sample of program stars which were observed on both the CTIO/IRS and IRTF/CSHELL systems. Here we assume the absolute velocity for the higher resolution IRTF/CSHELL data is correct; see text.

The mean radial velocities in our four fields are compared to the predictions for both dynamical models in Figure 4. All velocities in this figure are for the *galactocentric* reference frame. The axisymmetric prediction is again along a slit whose position angle is about 55° from the major axis. The dotted line in this figure is an estimate of the projected velocity as a function of $l$ assuming $V_r$ (-$l$) = $-V_r$ ($l$), which is a good approximation (Zhao, private communication). A detailed map of $V_r(l, b)$ for the bar model (Zhao 1994) shows essentially cylindrical rotation (the projected velocity is constant with latitude) with a linear dependence on longitude for the region of the inner Galaxy shown here.

The mean velocity in field 4 is significantly more negative than either model prediction. Part of this discrepancy may be due to the adopted correction to the galactocentric frame. A linear fit to the four data points suggests that the mean galactocentric velocity is $-15$ km s$^{-1}$ at $l = 0°$. This is nearly the same amount as the difference between the IRTF/CSHELL and CTIO/IRS velocities for a sub-set of stars measured on both systems (§3.1). If we correct for this offset by adding 12 km s$^{-1}$ to the mean velocities in fields 1,2 and 4 (they were measured on the CTIO/IRS system), then the mean velocities (open triangles in Figure 5) would match the Zhao bar model quite well. The results are also consistent with the Kent (1992) model.

## 4.3.  Velocity Distributions

The axisymmetric model and bar model both predict distributions of observed velocities which are quite similar to a Gaussian distribution for certain fields. Kuijken (1994) has derived the phase space distribution function for the Kent (1992) model. He finds that the predicted distribution of observed velocities towards Baade's window is well approximated by a Gaussian distribution. The detailed distribution function, which Kuijken derives numerically, fits the Baade's window M giant kinematics (Sharples et al. 1990). The Zhao (1994) model predicts an observed velocity distribution which is nearly Gaussian for $l$ near zero degrees but which can become strongly asymmetric at larger $l$ ($\sim 7$ °).

The observed *heliocentric* distribution of velocities for each field is shown in Figure 5. Kuijken has kindly provided us with predictions for our fields from the same model as described in Kuijken (1994); these are shown in Figure 5 for comparison. We have arbitrarily shifted the model predictions to the mean observed velocity; here we are only comparing the relative numbers of stars at each velocity. The mean velocities predicted by Kuijken's distribution function for our fields are similar to those of the Kent model which were discussed in the previous section. The Kuijken distributions were renormalized to give the same number of stars as observed in each field. The Kuijken (1994) model results are well fit by Gaussian distributions, but they have slightly lower dispersions than the Kent model ($\sim 10 \text{ km s}^{-1}$), and thus our observations. This is perhaps due to a truncation of high angular momentum stars which had to be made for computational reasons, or a slightly different disk model used than for the Kent model (Kuijken, private communication).

If we compare the observations to Gaussian distributions of the observed width and mean velocity (also shown in Figure 5), we find no statistical evidence that any of the observed distributions is different from Gaussian (by application of a Kolmogorov-Smirnov [KS] test to each field; the KS test does not depend on binning the data).

Zhao has also provided us with the predicted distribution in one of our fields (field 4, $l,b = -1°, 2°$) from his model (Figure 5). The model results were shifted and normalized as described above. Again, this distribution is well fit by a Gaussian and is quite similar to the Kuijken result. Zhao was forced to cut off the highest velocity stars due to storage requirements when producing the velocity distribution; this results in a reduced dispersion as computed from the distribution (the actual dynamical model calculation predicts a higher dispersion in this field as discussed above).

Due to the small number of stars in each field, we do not attempt to uniquely determine the form of the velocity distributions; however, the model results and KS tests suggest that a Gaussian is a good representation.

## 4.4. Other Models and Constraints

Our observed velocity dispersions and mean velocities appear to be marginally better fit by the Zhao (1994) bar model. This may be reassuring when we consider the photometric evidence that points to a non–axisymmetric stellar distribution. However, both the non–axisymmetric and axisymmetric models we have discussed here have similar axisymmetric components inside a minor axis distance of about 400 pc (3°), so a more stringent test of the axisymmetric *vs.* bar models in the innermost part of the bulge must await models with non-axisymmetric mass distributions there. Such models must still be consistent with the observed light distribution. Since it is very difficult to uniquely determine the unreddened surface brightness distribution and separate the bulge and disk in this region (Arendt et al. 1994), finding convincing non-axisymmetric models from the observed surface brightness distribution seems unlikely. Perhaps the M giant kinematics which we plan to obtain and/or a better understanding of the OH/IR star kinematics inside 100 pc will settle the question.

The Kent (1992) and Zhao (1994) models predict similar total masses for the bulge:

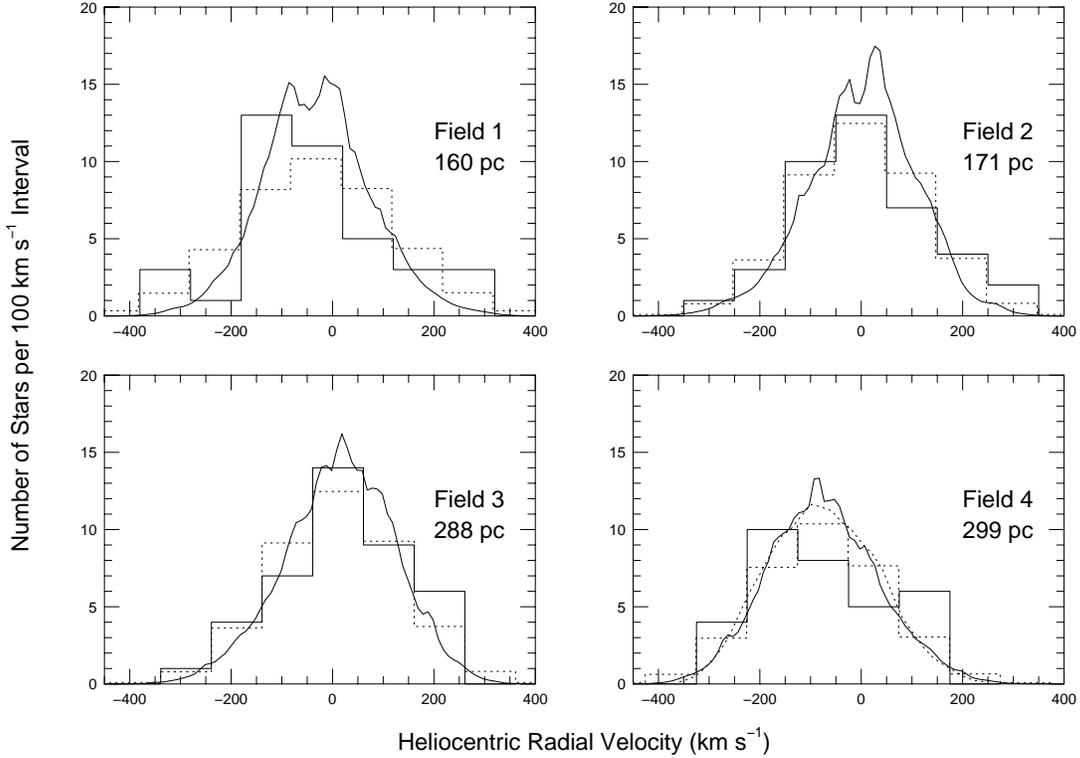

Fig. 5.— Comparison of observed and model velocity distributions. The model distributions have been arbitrarily shifted to the mean heliocentric velocity observed in each field. The solid histograms are for the four inner Galaxy fields discussed in this paper and Blum et al. (1994). The observations have been placed in 100 km s$^{-1}$ bins. The dashed histograms are Gaussian curves of the same observed full width and mean heliocentric velocity, binned to the same 100 km s$^{-1}$ as the observed data. The solid curves are predictions from the same distribution function as Kuijken (1994) which is based on the Kent (1992) model; see text. The dashed line for field 4 is the distribution from the model of Zhao (1994). Both model predictions are well represented by a Gaussian.

$1.8 \times 10^{10}$ M$_\odot$ *vs.* $2 \times 10^{10}$ M$_\odot$ for the Kent and Zhao models, respectively. The Kent (1992) value follows from direct integration of the model luminosity profile and derived mass-to-light ratio (Blum 1994). The total mass of the bulge may be even greater than these models imply. Blum (1994) has calculated the bulge mass using the tensor virial theorem with the global observed kinematics, the pattern speed determined by Binney et al. (1991), and the same Dwek et al. (1994) model employed by Zhao (1994). Blum (1994) finds the bulge mass could be as high as $2.8 \times 10^{10}$ M$_\odot$ if the bar major axis is aligned at $20°$ to our line of sight and the pattern speed is 81 km s$^{-1}$ kpc$^{-1}$. The uncertainty in the total bulge mass is represented by the differences in these three determinations. The Kent model applies to an axisymmetric system; Zhao's model is complex, and he has not yet fully explored the available parameter space for barred models; and Blum's estimate is sensitive to the bar orientation and pattern speed, neither of which is precisely known.

Self consistent bar models are difficult to construct due to their complexity and computational requirements. When a series of dynamical models can be constructed which account for the range of bar models which fit the light distribution, the wealth of kinematic data that now samples essentially all of the inner Galaxy may narrow the choice of models (and hence the bulge mass).

On the other hand, new types of observations may help to constrain the mass distribution. Three dimensional space motions of bulge stars offer the possibility of more stringent constraints on the determination of the inner Galaxy mass distribution (Spaenhauer, Jones, & Whitford 1992). For example, Zhao et al. (1994) have noted that by including the transverse components of velocity for the bulge stars one can definitively show whether or not the stars move on orbits which result from a non-axisymmetric potential. They analyzed a sample of K giants in Baade's window which have both radial velocities and proper motions. From these, they compute the off-diagonal terms to the velocity dispersion tensor and find that the observed velocity ellipsoid is not aligned with the spatial directions (line of sight and *l*) as it should be if the mass distribution were axisymmetric. Their result depends on breaking the sample of stars in to metal rich and metal poor subsamples. The observed "vertex deviation" is not present in the combined sample. Rich, Terndrup, & Sadler (1994) have three space motions for a much larger sample of bulge K giants. Analysis of this data set will more clearly demonstrate whether vertex deviations are present.

The bulge mass distribution may also be constrained by observations of microlensing events. Such events occur when a foreground object passes near in projection to a background star and "focuses" the background star light due to its gravitational field, creating a magnification of the background star which can be detected. Approximately 60 microlensing events have been detected by the two groups searching for them toward the bulge (The Optical Gravitational Lensing Experiment, Udalski et al. 1994; Massive Compact Halo Object collaboration, Bennett et al. 1994). The total number of events observed is better explained by a barred model of the bulge than an axisymmetric one (e.g. Han & Gould 1994). The total number of events observed so far is still small enough to allow for a range of bar models to fit the data (Han & Gould 1994).

## 5. SUMMARY

We have obtained radial velocities for approximately 120 M giant stars in three fields at projected radii between 150 and 300 pc. Combined with the data presented by Blum et al. (1994), we have a total of about 150 stars in four fields.

The mean velocities, velocity dispersions, and velocity distributions of the observed samples agree favorably with available axisymmetric (Kent 1992) and barred (Zhao 1994) dynamical models. Figures 3 and 4 suggest a slightly better representation of the observed kinematics by the bar model.

The velocity dispersion in our innermost field is 153 km s$^{-1}$ $\pm$ 17 km s$^{-1}$, 1 to 2 $\sigma$ ($\sim$ 15 $-$ 30 km s$^{-1}$) higher than either model prediction and higher than observed anywhere else in the inner Galaxy. This means that we have yet to observe the dispersion begin decreasing toward smaller radii as we might expect if the bulge and GC populations are to join smoothly together.

We are indebted to Stephen Kent, Konrad Kuijken, Jerry Sellwood, and HongSheng Zhao for making available to us model predictions as well as for helpful discussions. We thank R. Pogge for use of his spectral line analysis program, LINER. This work was supported by a grant from National Science Foundation (AST - 9115236).

Table 1: **M Giant Fields and Mean Properties**

| Field | $l^\circ$ | $b^\circ$ | $R^a$(pc) | $CO_{sp}{}^b$ | $CO_{ph}{}^c$ | $(J-K)_\circ{}^d$ | $A_K{}^e$ | $\overline{V_r}{}^f$(km s$^{-1}$) | $\sigma$ (km s$^{-1}$) |
|---|---|---|---|---|---|---|---|---|---|
| 1 | $-0.59$ | 0.98 | 160 | 0.30 | 0.21 | 1.05 | 1.06 | $-31 \pm 28$ | $153 \pm 17$ |
| 2 | 0.85 | $-0.88$ | 171 | $0.32^g$ | 0.23 | 1.10 | 0.79 | $0 \pm 23$ | $128 \pm 14$ |
| 3 | 1.21 | $-1.67$ | 288 | 0.29 | .20 | 1.05 | 0.29 | $14 \pm 23$ | $128 \pm 14$ |
| $4^h$ | $-1.14$ | 1.81 | 299 | 0.33 | 0.24 | 1.10 | 0.63 | $-75 \pm 24$ | $127 \pm 17$ |

[a] Projected distance from the Galactic center, assuming $R_\circ = 8$ kpc

[b] Observed mean spectroscopic CO strength

[c] Photometric CO index converted from spectroscopic CO strength; see text

[d] intrinsic color based on CO index vs. intrinsic $J-K$ of FW; see text.

[e] Derived using the extinction law of Mathis (1990) and intrinsic color based on CO strength; see text. Uncertainty in $A_K$ is $\sim 0.01$ mag from photometric uncertainty alone.

[f] Heliocentric radial velocity

[g] CO converted to lower resolution measure; see text.

[h] CO and observed colors taken from Blum et al. (1994), but intrinsic colors and extinction derived here

Table 2.   M Giant Observed Magnitudes and Radial Velocities

| Star | $\Delta\alpha('')$[a] | $\Delta\delta('')$[a] | $V_r$[b]$(km\ s^{-1})$ | $K$[c] | $J-K$[c] | CO[d] |
|------|------|------|------|------|------|------|
| 1−1 | 123 | 19 | −63 | 8.04 | 2.59 | 0.24 |
| 1−2 | −95 | −15 | 113 | 8.06 | 2.76 | 0.32 |
| 1−3 | −83 | 20 | −69 | 8.20 | 2.66 | 0.32 |
| 1−4 | −67 | −51 | −282 | 8.24 | 2.89 | 0.35 |
| 1−5 | −66 | −161 | 26 | 8.26 | 2.96 | 0.28 |
| 1−6 | −17 | 17 | 80 | 8.29 | 2.66 | 0.30 |
| 1−7 | 90 | 179 | 19 | 8.32 | 2.52 | 0.28 |
| 1−8 | 74 | 42 | −92 | 8.38 | 2.82 | 0.27 |
| 1−9 | −92 | −69 | −85 | 8.39 | 2.92 | 0.35 |
| 1−10 | −40 | 67 | −88 | 8.40 | 2.71 | 0.32 |
| 1−11 | −158 | 86 | −142 | 8.42 | 2.56 | 0.24 |
| 1−12 | −96 | −89 | 142 | 8.42 | 2.77 | 0.34 |
| 1−13 | −14 | −117 | −71 | 8.46 | 2.81 | 0.29 |
| 1−14 | −166 | −104 | −224 | 8.49 | 2.77 | 0.30 |
| 1−15 | −85 | 117 | 272 | 8.50 | 2.68 | 0.19 |
| 1−16 | −93 | 145 | −163 | 8.50 | 2.61 | 0.22 |
| 1−17 | −163 | −119 | −6 | 8.53 | 2.75 | 0.29 |
| 1−18 | −104 | −41 | −363 | 8.53 | 2.50 | 0.36 |
| 1−19 | 78 | 103 | 210 | 8.59 | 2.59 | 0.28 |
| 1−20 | −138 | 111 | 116 | 8.63 | 2.62 | 0.32 |
| 1−21 | −85 | −112 | −87 | 8.68 | 2.73 | 0.28 |
| 1−22 | −82 | −18 | −84 | 8.70 | 2.51 | 0.27 |
| 1−23 | 132 | 63 | −28 | 8.73 | 2.67 | 0.30 |
| 1−24 | 10 | 95 | 9 | 8.76 | 2.73 | 0.24 |
| 1−25 | 31 | −103 | −98 | 8.78 | 2.71 | 0.27 |
| 1−26 | −120 | −100 | −35 | 8.80 | 2.71 | 0.34 |
| 1−27 | 80 | −134 | 198 | 8.80 | 2.97 | 0.30 |
| 1−28 | −2 | −167 | −175 | 8.80 | 2.93 | 0.34 |
| 1−29 | 34 | −112 | −93 | 8.82 | 2.86 | 0.32 |
| 1−30 | −140 | 118 | −17 | 8.82 | 2.65 | 0.32 |
| 1−31 | 186 | −23 | 270 | 8.83 | 3.17 | 0.25 |
| 1−32 | 49 | −90 | −176 | 8.86 | 2.86 | 0.30 |
| 1−33 | 184 | −97 | −316 | 8.94 | 3.14 | 0.32 |
| 1−34 | −29 | −86 | 281 | 8.96 | 2.67 | 0.32 |

Table 2—Continued

| Star | $\Delta\alpha('')^{\mathrm{a}}$ | $\Delta\delta('')^{\mathrm{a}}$ | $V_r{}^{\mathrm{b}}(\mathrm{km\ s^{-1}})$ | $K^{\mathrm{c}}$ | $J-K^{\mathrm{c}}$ | $CO^{\mathrm{d}}$ |
|------|------|------|------|------|------|------|
| 1−35 | −72 | −115 | −104 | 8.98 | 2.51 | 0.17 |
| 1−36 | −151 | −134 | 58 | 8.99 | 2.78 | 0.26 |
| 1−37 | −119 | 149 | 5 | 9.00 | 2.78 | 0.40 |
| 1−38 | 69 | −85 | −28 | 9.02 | 2.83 | 0.39 |
| 1−39 | −106 | 57 | −100 | 9.03 | 2.55 | 0.21 |
| 2− 1 | −36 | −130 | 58$^{\mathrm{e}}$ | 7.86 | 2.26 | 0.34 |
| 2− 2 | −156 | −73 | −23 | 7.88 | 2.80 | 0.48 |
| 2− 3 | 146 | 87 | −192 | 7.91 | 2.26 | 0.58 |
| 2− 4 | 147 | 108 | −77 | 7.91 | 2.19 | 0.48 |
| 2− 5 | −150 | −46 | −268 | 7.91 | 2.57 | 0.43 |
| 2− 6 | −54 | −67 | 252$^{\mathrm{f}}$ | 7.94 | 2.23 | $\cdots$ |
| 2− 7 | −83 | 126 | 24 | 7.99 | 2.69 | 0.55 |
| 2− 8 | 139 | −158 | −108 | 8.01 | 2.36 | 0.48 |
| 2− 9 | 85 | −32 | −56$^{\mathrm{e}}$ | 8.01 | 2.18 | 0.11 |
| 2−10 | −4 | −18 | −232 | 8.02 | 2.33 | 0.51 |
| 2−11 | −88 | 92 | 68 | 8.04 | 2.34 | 0.53 |
| 2−12 | −41 | −87 | −136 | 8.11 | 2.18 | 0.57 |
| 2−13 | −33 | −53 | 158 | 8.13 | 2.31 | 0.51 |
| 2−14 | −15 | 153 | 237$^{\mathrm{f}}$ | 8.15 | 2.27 | $\cdots$ |
| 2−15 | 1 | 27 | −30 | 8.17 | 2.32 | 0.56 |
| 2−16 | −136 | 97 | −20 | 8.18 | 2.41 | 0.46 |
| 2−17 | −121 | 167 | −18 | 8.22 | 2.34 | 0.41 |
| 2−18 | 138 | −98 | 216$^{\mathrm{f}}$ | 8.22 | 2.21 | $\cdots$ |
| 2−19 | 2 | 66 | −172 | 8.23 | 2.23 | 0.52 |
| 2−20 | −35 | 126 | 52 | 8.26 | 2.21 | 0.50 |
| 2−21 | 40 | 187 | −64 | 8.30 | 2.54 | 0.53 |
| 2−22 | −176 | 35 | 53 | 8.32 | 2.55 | 0.50 |
| 2−23 | −62 | 117 | 278$^{\mathrm{f}}$ | 8.36 | 2.50 | $\cdots$ |
| 2−24 | 15 | 153 | −82 | 8.37 | 2.39 | 0.53 |
| 2−25 | 70 | −97 | 113 | 8.41 | 2.18 | 0.52 |
| 2−26 | −62 | −106 | −109 | 8.42 | 2.41 | 0.52 |
| 2−27 | −136 | 181 | −8 | 8.45 | 2.36 | 0.50 |
| 2−28 | −139 | −93 | −105 | 8.47 | 2.50 | 0.52 |
| 2−29 | −165 | −97 | −130 | 8.53 | 2.19 | 0.41 |

Table 2—Continued

| Star | $\Delta\alpha('')$[a] | $\Delta\delta('')$[a] | $V_r$[b]$(km\ s^{-1})$ | $K$[c] | $J-K$[c] | CO[d] |
|------|------|------|------|------|------|------|
| 2−30 | −144 | 133 | 74 | 8.55 | 2.36 | 0.48 |
| 2−31 | −92 | 131 | 29 | 8.57 | 2.43 | 0.56 |
| 2−32 | 57 | −74 | 182[f] | 8.58 | 2.23 | ⋯ |
| 2−33 | 25 | −26 | 22 | 8.59 | 2.20 | 0.48 |
| 2−34 | 145 | −78 | −83 | 8.61 | 2.34 | 0.52 |
| 2−35 | −90 | 192 | 22 | 8.62 | 2.31 | 0.52 |
| 2−36 | −40 | 63 | 4 | 8.63 | 2.37 | 0.48 |
| 2−37 | 137 | 135 | 20 | 8.71 | 2.37 | 0.44 |
| 2−38 | −143 | 112 | 48 | 8.72 | 2.22 | 0.42 |
| 2−39 | −27 | 40 | 1 | 8.77 | 2.20 | 0.48 |
| 3−1 | −52 | 140 | 76 | 7.38 | 1.68 | 0.32 |
| 3−2 | −102 | −34 | 182 | 7.44 | 1.67 | 0.26 |
| 3−3 | 139 | 31 | −152 | 7.57 | 1.60 | 0.32 |
| 3−4 | 130 | 10 | −104 | 7.66 | 1.69 | 0.32 |
| 3−5 | 62 | −131 | 210 | 7.67 | 1.64 | 0.32 |
| 3−6 | 3 | −165 | −183 | 7.68 | 1.73 | 0.27 |
| 3−7 | 60 | −164 | 140 | 7.85 | 1.58 | 0.30 |
| 3−8 | −146 | −23 | −18 | 7.92 | 1.53 | 0.25 |
| 3−9 | −94 | −45 | −50 | 7.97 | 1.62 | 0.32 |
| 3−10 | −18 | −156 | −17 | 8.00 | 1.53 | 0.29 |
| 3−11 | 11 | 54 | −10 | 8.15 | 1.51 | 0.29 |
| 3−12 | 76 | 128 | 58 | 8.16 | 1.50 | 0.20 |
| 3−13 | 17 | 26 | 254 | 8.17 | 1.53 | 0.30 |
| 3−14 | −154 | 160 | 56 | 8.21 | 1.61 | 0.26 |
| 3−15 | −15 | 166 | 0 | 8.26 | 1.73 | 0.36 |
| 3−16 | 107 | −66 | 18 | 8.27 | 1.65 | 0.33 |
| 3−17 | −105 | −175 | −134 | 8.28 | 1.44 | 0.35 |
| 3−18 | −165 | −110 | 132 | 8.35 | 1.49 | 0.34 |
| 3−19 | 5 | −11 | 18 | 8.36 | 1.66 | 0.25 |
| 3−20 | −114 | −159 | −187 | 8.38 | 1.39 | 0.27 |
| 3−21 | −57 | 101 | −128 | 8.43 | 1.65 | 0.31 |
| 3−22 | 53 | −174 | 220 | 8.46 | 1.50 | 0.35 |
| 3−23 | −118 | 169 | 10 | 8.47 | 1.57 | 0.25 |
| 3−24 | 22 | 12 | 78 | 8.48 | 1.40 | 0.30 |

[a]Offsets are from SAO 185604 ($\alpha(1950)=17^h\ 37^m\ 13.48^s$, $\delta(1950)=-28°\ 53'\ 50.0''$), SAO 185837 ($\alpha(1950)=17^h\ 47^m\ 53.3^s$, $\delta(1950)=-28°\ 38'46.8''$), and SAO 185927 ($\alpha(1950)=17^h\ 51^m\ 48.83^s$, $\delta(1950)=-28°\ 44'\ 53.7''$) for fields 1,2, and 3, respectively; see text. The offsets are accurate to $\sim 1''-2''$.

[b]Heliocentric radial velocity. Velocities are uncertain by approximately $\pm$ 15 km s$^{-1}$; see text.

[c]Observed magnitudes, uncorrected for interstellar extinction. The $K$ and $J$ magnitudes have photometric uncertainties of 0.07 and 0.05 mag, respectively.

[d]The CO strength is defined as $1 - F_{CO}/F_{con}$, where $F_{CO}$ is the flux in a band just longward of the bandhead and $F_{con}$ is the flux in a similar band just shortward of the bandhead. The band width was 0.005 $\mu$m for fields 1 and 3 and 0.001 $\mu$m for field 2.

[e]Heliocentric velocity and CO strength measured on CTIO/IRS system.

[f]Heliocentric velocity estimated from position of the bandhead, not by cross–correlation, and no measured CO strength because the large velocity shifted part of the CO band off the array.

Table 2—Continued

| Star | $\Delta\alpha('')^a$ | $\Delta\delta('')^a$ | $V_r{}^b(\text{km s}^{-1})$ | $K^c$ | $J - K^c$ | $CO^d$ |
|------|------|------|------|------|------|------|
| 3−25 | −181 | −98 | −38 | 8.52 | 1.42 | 0.24 |
| 3−26 | 146 | −92 | −136 | 8.54 | 1.44 | 0.27 |
| 3−27 | 156 | −7 | −102 | 8.55 | 1.41 | 0.27 |
| 3−28 | −153 | 103 | 59 | 8.55 | 1.52 | 0.26 |
| 3−29 | −27 | −103 | 180 | 8.56 | 1.49 | 0.32 |
| 3−30 | 42 | −5 | −20 | 8.56 | 1.59 | 0.33 |
| 3−31 | −122 | −93 | −4 | 8.59 | 1.42 | 0.30 |
| 3−32 | −35 | −116 | 66 | 8.60 | 1.45 | 0.31 |
| 3−33 | 29 | −110 | −282 | 8.65 | 1.43 | 0.28 |
| 3−34 | −57 | −186 | 140 | 8.68 | 1.50 | 0.34 |
| 3−35 | −166 | −135 | 68 | 8.72 | 1.31 | 0.26 |
| 3−36 | −63 | −54 | −168 | 8.75 | 1.47 | 0.32 |
| 3−37 | 89 | −4 | 85 | 8.78 | 1.32 | 0.21 |
| 3−38 | 22 | 63 | 166 | 8.80 | 1.44 | 0.30 |
| 3−39 | 46 | 158 | 146 | 8.81 | 1.35 | 0.20 |
| 3−40 | −155 | −104 | −87 | 8.81 | 1.35 | 0.21 |